\documentclass[letterpaper, 10 pt, conference]{ieeeconf}  
\usepackage{graphicx}
\usepackage{float}
\usepackage{hyperref}
\usepackage{subfig}
\usepackage[dvipsnames]{xcolor}

\usepackage[backend=biber,style=numeric,citestyle=numeric,,sorting=none,maxnames=50]{biblatex}
\IEEEoverridecommandlockouts                           
\title{\LARGE \bf
AIris: An AI-powered Wearable Assistive Device\\for the Visually Impaired
}

\author{Dionysia Danai Brilli $^{1, * }$, Evangelos Georgaras$^{1}$, Stefania Tsilivaki$^{1}$, Nikos Melanitis $^{1}$, Konstantina Nikita $^{1}$
\thanks{$^{1}$ Electrical \& Computer Engineering, National Technical University of Athens }%
\thanks{$^{*}$ Danai Brilli:  {\tt\small danaibrilli@gmail.com}}%
}
\begin{document}

\maketitle
\thispagestyle{empty}
\pagestyle{empty}

\begin{abstract}
    Assistive technologies for the visually impaired have evolved to facilitate interaction with a complex and dynamic world. In this paper, we introduce AIris, an AI-powered wearable device that provides environmental awareness and interaction capabilities to visually impaired users. AIris combines a sophisticated camera mounted on eyewear with a natural language processing interface, enabling users to receive real-time auditory descriptions of their surroundings. We have created a functional prototype system that operates effectively in real-world conditions.  AIris demonstrates the ability to accurately identify objects and interpret scenes, providing users with a sense of spatial awareness previously unattainable with traditional assistive devices. The system is designed to be cost-effective and user-friendly, supporting general and specialized tasks: face recognition, scene description, text reading, object recognition, money counting, note-taking, and barcode scanning. AIris marks a transformative step, bringing AI enhancements to assistive technology, enabling rich interactions with a human-like feel.
    \end{abstract}

\begin{keywords}
    Assistive Technology, Visually Impaired, Computer Vision, Deep Learning, Wearable Device, Wearable Robotics, Human Performance Augmentation, Virtual Reality and Interfaces
\end{keywords}

\section{INTRODUCTION}
    Artificial Intelligence has become a huge part of our everyday lives, finding its way into many different aspects and industries. AI has changed the way we communicate, shop, entertain ourselves, to even the way healthcare works. Advancements in medical imaging and AI have revolutionized medical practice, from early disease detection and diagnosis to personalized medicine. AI has made our world a lot more accessible by enabling smart environments and offering a better quality of life to people with disabilities \cite{PintoCoelho2023}.

    According to the World Health Organization, there are 253 million people estimated to be visually impaired and 36 million people totally blind in the world \cite{Bourne2017}. Blind people face difficulties navigating the outdoors, engaging in social activities, browsing the internet, or even everyday activities, like currency denomination or house chores. So, visual impairment poses a significant challenge to the perception and interpretation of environmental stimuli, showcasing the need for auxiliary technological aids. Assistive technology for the visually impaired has seen great advancements in recent years, including electronic travel aids, like Smart Cane \cite{SmartCane}, Bbeep \cite{Kayukawa2019} or Shoe \cite{shoe}, Orientation Aids like Smart

    \begin{figure}[H]
        \centering
        {
        \includegraphics[scale=0.45]{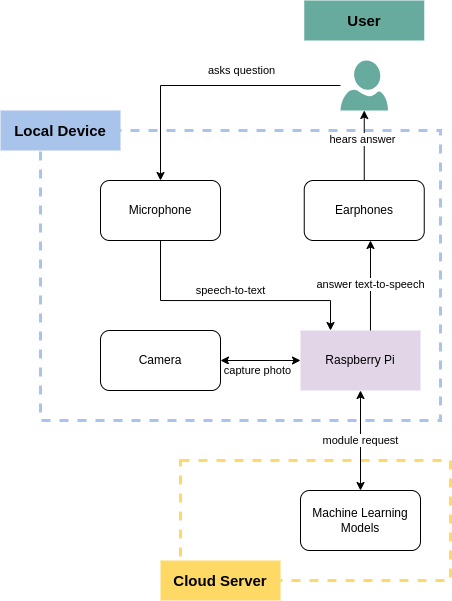}
        }
        \caption{AIris architecture overview: User is wearing the device, including earphones, camera and Raspberry Pi. The user asks a question, which Raspberry Pi converts to text and using Natural Language Processing it selects which module should be evoked. The camera input is transmitted to the server, where the ML model inference happens. The results are then sent back to the Raspberry Pi, where they are formatted into a sentence, converted from text to speech, and finally communicated to the user through the earphones.}
        \label{figurelabel}
    \end{figure}
    \noindent
     Walker \cite{SmartWalker} and even Position Locator Devices \cite{Manjari2020}. However, current assistive devices offer limited functionality, failing to provide comprehensive spatial awareness and independence \cite{Manjari2020}.

    \begin{figure*}[t]
        \centering
        \includegraphics[width=0.9\textwidth]{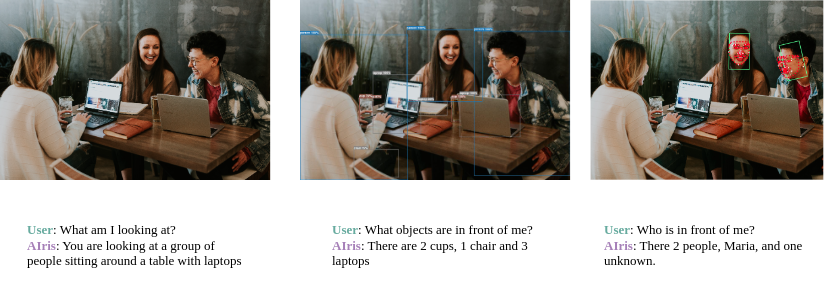}
        \caption{An example of how the Scene Description, Object Recognition and Face Recognition Modules interact with the user and the captured data. Photo by Brooke Cagle on Unsplash \cite{unsplashPhotoBrooke}}
        \label{fig:img1}
    \end{figure*}
    
    In this study, we develop AIris, an assistive system for the visually impaired with AI-enhanced functionalities. AIris uses state-of-the-art machine learning models to facilitate environmental understanding through auditory feedback. We design and build a prototype of the assistive device, by designing a custom pair of 3D-printed glasses equipped with a camera slot, earphones, and a microcomputer. We develop a fully operational prototype system that can be used in varying real-life scenarios and daily activities. Using computer vision techniques and advanced pre-trained neural networks, it analyzes camera data to provide comprehensive feedback about face detection, text reading, object recognition, money counting, and many more.  AIris functionalities aim to make the environment more accessible and give visually impaired people greater independence in their everyday activities.

\section{METHODS}
    \subsection{System Overview}
        The hardware infrastructure of AIris is designed to fulfill the needs of visually impaired users. We use a distributed architecture,  capturing local information and transmitting to a central server for computationally-intensive processing (Fig.~\ref{figurelabel}). The device's primary sensor is a camera, centrally positioned on the eyewear, capturing high-resolution visual data from the user's perspective. This visual information is initially processed by a Raspberry Pi microcomputer, serving as a local computational node.

        The microcomputer then securely transmits the data to a centralized server, where intensive image processing and analysis are conducted. The server, leveraging superior computational resources, applies advanced machine learning models for tasks such as object detection, scene description, and text extraction. Upon completing the analysis, the server dispatches the processed results back to the Raspberry Pi.

        On receipt of the server's data, the Raspberry Pi undertakes the final assembly of the response using a natural language processing (NLP) engine. The NLP engine is responsible for translating analytical data into intelligible auditory feedback, which is then communicated to the user via the earphones. This processing approach —local data acquisition and remote analysis— ensures that AIris provides rapid and accurate environmental descriptions, enhancing the user's spatial awareness with minimal latency.

    \subsection{System Capabilities}
        \begin{figure*}
            \centering
            \includegraphics[width=\textwidth]{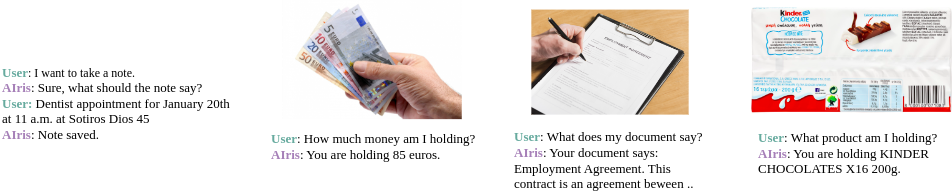}
            \caption{An example of how the Note Taking, Counting Money, Text Reading and Barcode Scanning Modules interact with the user and the captured data. Photo in the middle by kstudio on Freepik \cite{freepikFreePhoto}}
            \label{fig:img2}
        \end{figure*}

        The AIris system is built from several key modules, each performing a distinct task.

        \subsubsection{Face Recognition Module}
            This module detects and recognizes human faces within the camera's visual field (Fig.~\ref{fig:img1}) \cite{FaceRecognition}. It checks recognized faces against a user-defined dynamic database and tells the user who the person is if known, otherwise, the users can choose to add the new faces to the system. This is essential for personalizing the user experience, as it provides the ability to remember and identify acquaintances, family members, and friends.
            To safeguard user privacy, we exclusively store facial landmark embeddings, eliminating the need to retain the full image. Due to the low computational requirements and to ensure low latency, this module is implemented on the Raspberry Pi, running locally.

        \subsubsection{Scene Description Module}
            The scene description module provides users with a detailed verbal representation of their immediate surroundings (Fig.~\ref{fig:img1}) \cite{Max}. It essentially acts as the eyes for the visually impaired, translating visual cues into auditory information that can be used to understand complex environments. This module employs advanced image segmentation alongside object detection algorithms to dissect the visual scene into identifiable components, such as obstacles, pathways, furniture, and other individuals and combine that information to describe the scene \cite{Max}.

        \subsubsection{Text Reading Module}
            The text reading module is a critical component of the AIris system, serving as a bridge between the visually impaired and the vast world of written communication. Utilizing state-of-the-art optical character recognition (OCR) technology,  based on Google's Tesseract, this module scans the visual field for textual content, interprets it, and converts it into spoken words \cite{Tesseract}. This module enables the visually impaired to access the vast textual information in labels, instructions, signs and documents.

        \subsubsection{Object Recognition Module}
            The object recognition module harnesses a pre-trained version of Yolo  \cite{Yolo} to identify and classify objects within the camera's field of view (Fig.~\ref{fig:img1}). The module is particularly significant as it allows users to interact with their environment in a more meaningful way, recognizing items that are used in daily life, from personal belongings to potential hazards.

        \subsubsection{Counting Money}
             This module employs a combination of custom image processing and pattern recognition algorithms to differentiate and count banknotes and coins (Fig.~\ref{fig:img2}). Its importance is underscored by the critical need for financial independence and confidence to conduct transactions without assistance.

        \subsubsection{Note Taking Module}
            This module utilizes advanced speech recognition technology to transcribe spoken words, save them, and access them later as needed (Fig.~\ref{fig:img2}). The significance of this module emerges from its facilitation of personal organization and memory assistance, which are essential for maintaining independence.

        \subsubsection{Barcode Scanning Module}
            The barcode scanning module swiftly interprets barcode data and informs the user about the product name, price, and generally any information available in public databases (Fig.~\ref{fig:img2}).

    \subsection{User Interaction}
        \subsubsection{Initial Setup \& Activation}
            The user interaction with AIris is initiated by a simple voice command, like most virtual assistants nowadays. When first activated, the system runs a quick device status check, including battery level, network connectivity etc.
            
        \subsubsection{Voice Commands Interface}
            Upon system activation, the user is invited to communicate with AIris using natural language. This interaction is facilitated by an advanced NLP engine, Google Speech \cite{GoogleSpeech}. Users issue commands or queries in natural language, such as requesting a description of their environment, identifying an object, or reading visible text. The NLP engine analyzes the speech, mapping the spoken phrases, using defined keywords, to predefined operations that the AIris system can execute.

        \begin{figure*}[htbp]
          \centering
          \subfloat[First Sketches]{\includegraphics[width=0.3\textwidth]{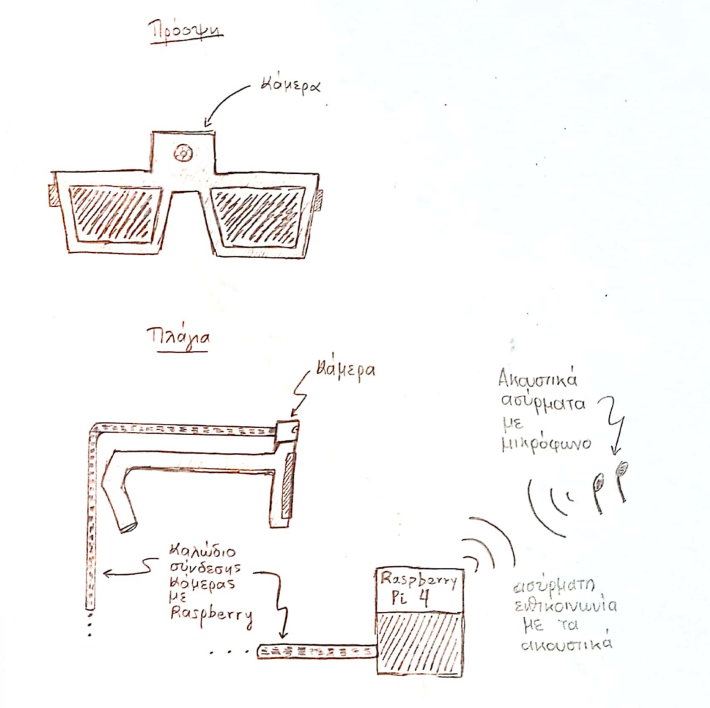} \label{fig:a }}
          \hfill 
          \subfloat[Computer Aided Design]{\includegraphics[width=0.3\linewidth]{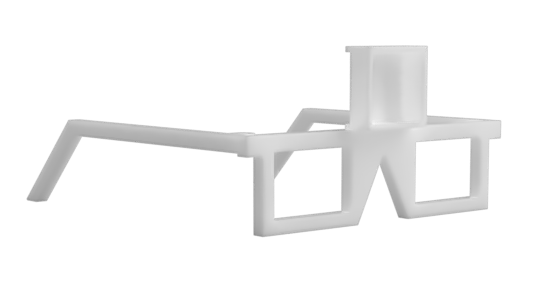}\label{fig:b}}
          \hfill 
          \subfloat[Fully-functional prorotype]{\includegraphics[width=0.3\linewidth]{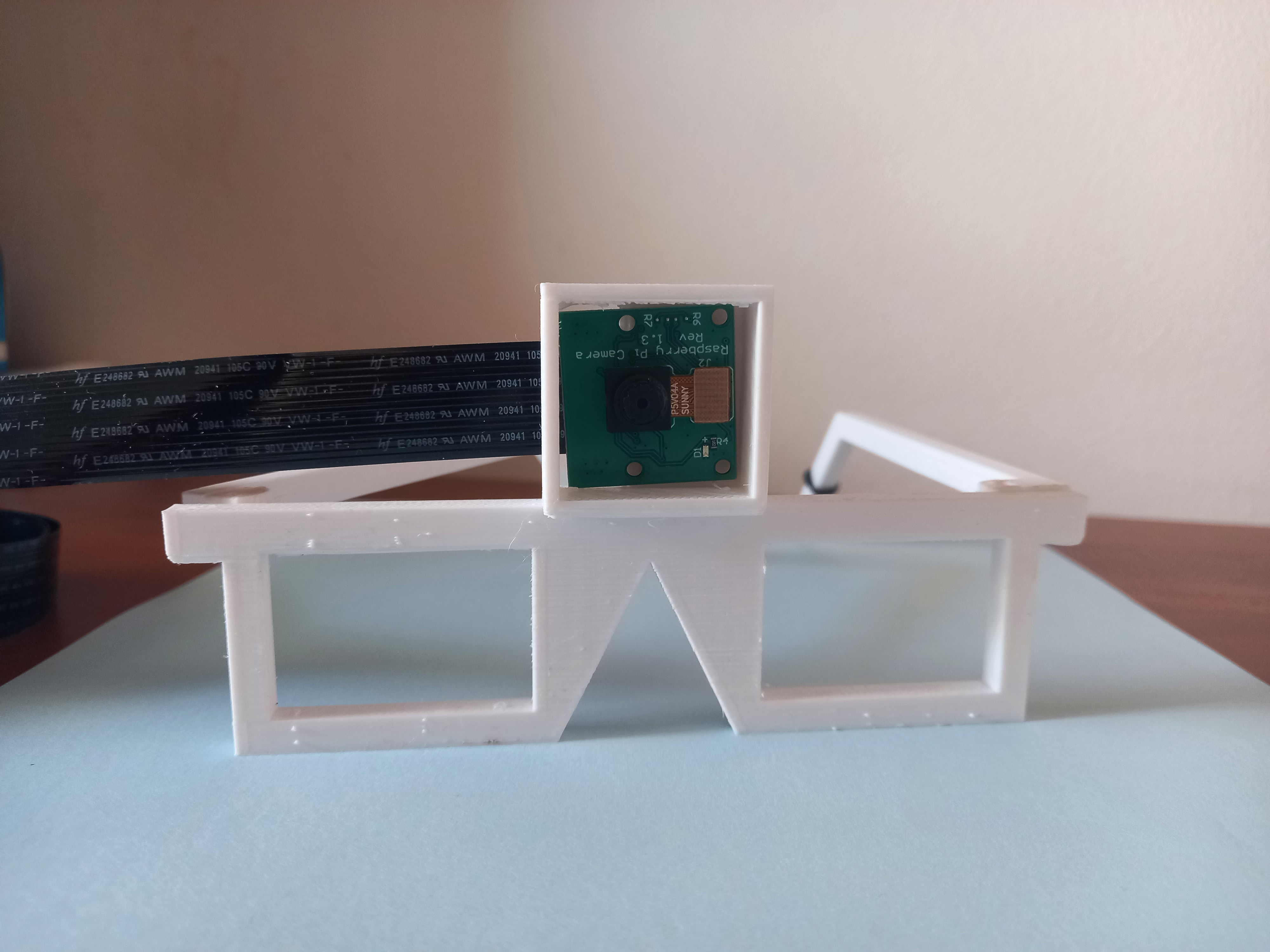}\label{fig:c}}
          \caption{AIris Design Process: This figure shows from left to right the evolution of the hardware for AIris. At first, we sketched out the basic idea and how the components would interact. Then we moved on to designing the eyewear in a Computer Aided Design (CAD) Program, focusing on keeping it light-weight, simple and user-friendly. Finally, we 3D printed our eyewear and assembled the whole prototype.}
          \label{fig:img3}
        \end{figure*}

        \subsubsection{Custom 3D Printed Eye Wear Design}
            The eyewear is designed meticulously in order to avoid obstructing the field of view of low-vision individuals. The objective was to develop a lightweight, comfortable frame capable of supporting a camera mount. The camera mount was engineered to position the camera at the center of the glasses, mirroring the natural human line of sight. Various prototypes were 3D printed to test the balance, fit, and comfort of the eyewear, with adjustments made to the CAD models in response to user feedback. The design process is illustrated through the iterations in Fig.~\ref{fig:img3}. 

        \subsubsection{Hardware Integration}
            In the prototypical stage of AIris, the thoughtful integration of hardware components has been a critical focus to ensure functionality while maintaining user comfort. The camera is strategically mounted at the front center of the eyewear. It is connected via a flexible, lightweight cable to the Raspberry Pi microcomputer, which is conveniently placed within the user's pocket. This arrangement keeps the eyewear as light as possible, reducing the strain on the user’s nose and ears during prolonged use.

            The Raspberry Pi 4B microcomputer, selected for its compact design and powerful processing capabilities, processes visual data at high speeds with minimal power consumption. To ensure extended mobility, the microcomputer utilizes a slim 20,000mAh power bank, providing over 6 hours of uninterrupted use.

            The Bluetooth earphones are paired seamlessly with the system, while also enabling ambient mode, enhancing the user's surroundings' sounds so as to not reduce their hearing.

        \subsubsection{Local Software \& Raspberry Pi}
            The local software running on the Raspberry Pi within the AIris system is chosen with respect to efficacy and performance \cite{RaspberryPi}. The first step of the user interaction is the speech-to-text conversion, facilitated by Google's Speech Recognition Engine \cite{GoogleSpeech}. This service translates spoken language into textual commands with very high accuracy, even in noisy environments like a supermarket, or a cafeteria, while also maintaining minimal latency.

            The processed commands are analyzed by a decision-making algorithm, which determines the appropriate module to invoke based on the context of the user's request. The decision-making algorithm is currently relying on a pre-determined set of keywords and word embeddings similarity.

            For facial recognition, the system employs face\_recognition, an advanced library known for its precision in detecting and identifying human faces, using dlib's version of  the ResNet-34 \cite{FaceRecognition}. A frame is isolated from the camera and the regions containing a face are identified. The locations of a face's landmarks are then recorded, fed into a neural network and encoded into a latent space of $128$ encodings. These encodings are then compared to a pre-existing database of known individuals, all while maintaining real-time processing speeds.

            When textual content needs to be extracted from the environment, the system leverages pytesseract, a Python wrapper for Google's Tesseract-OCR Engine \cite{Tesseract}. Pytesseract is adept at recognizing text within a wide variety of lighting conditions and font styles.

            The counting money module utilizes computer vision filters and optical character recognition (OCR) for identification and counting of currency, drawing inspiration from the work detailed in \cite{money}. Upon receiving the camera input, the server applies filters and thresholding to enhance the features of the banknotes or coins. Following this enhancement, the pytesseract OCR engine is employed to read the denominations. The module is adept at handling various currencies, orientations, and lighting conditions, providing users with an accurate count and denomination of their money. 

            The note-taking module is a significant feature that enhances the user's ability to record and organize information. This module operates by converting spoken words into text using Google's Speech Recognition engine \cite{GoogleSpeech}. Once the speech is transcribed, the system categorizes the text based on user-defined categories such as reminders, notes, or lists. The categorized notes are then saved into text files which can be easily retrieved and reviewed by the user.

            To achieve efficient and accurate barcode scanning, this module leverages Pyzbar, a python library known for decoding barcodes and QR codes\cite{PyZBar}. When the user points the AIris camera at a barcode, Pyzbar processes the image to detect and decode the barcode. Given the varying qualities of barcodes and the diverse visual conditions, the server employs image processing techniques to enhance the barcode's features. These preprocessing steps include adjusting brightness and contrast, sharpening, and scaling to ensure that Pyzbar can accurately decode the information even in less-than-ideal circumstances.

            Finally, the Raspberry Pi utilizes a text-to-speech library, pyttsx3, and the espeak engine to convert the textual information back into speech \cite{pyttsx3}, \cite{espeak}. This final step is crucial as it provides the user with a natural-sounding voice that conveys the results of the various AI modules.

        \subsubsection{Server Software}
            The server-side software of AIris runs on DeepDetect, an Open-Source Deep Learning platform, running pre-trained state-of-the-art machine learning models, each dedicated to a specific module.  The server side, thus, performs intricate analyses and computations that exceed the processing capabilities of the Raspberry Pi.

            For the scene description module, we deploy MAX (Model Asset eXchange), an Image Caption Generator \cite{Max}. This model generates captions from a fixed vocabulary that describe the contents of images in the COCO Dataset \cite{COCO}. The model consists of an encoder model - a deep convolutional net using the Inception-v3 architecture trained on ImageNet-2012 data - and a decoder model \cite{Inception}, \cite{imagenet}. The input to the model is an image, and the output is a sentence describing the image content.

            For Object Recognition, we use YOLO \cite{Yolo}. When the user requests object identification, the server processes the visual input to pinpoint and identify objects, count each class occurrence, and relay this information back to the Raspberry Pi for the final answer formation.
            
\section{RESULTS \& DISCUSSION}
    AIris is a fully-fledged solution that captures data, processes it, and provides users with cues. AIris modules include face recognition, scene description, text reading, object recognition, money counting, note-taking and barcode scanning. AIris uses deep learning models and machine learning algorithms for efficient processing of visual data.

    AIris users  may:
    \begin{itemize}
        \item rapidly identify familiar faces and save new ones
        \item get detailed description of their surroundings, to better navigate through environments
        \item read digital \& printed text through audio output, making every menu, sign or document accessible 
        \item identify and locate objects around them
        \item get auditory description of the currency they hold
        \item record verbal notes, aiding in personal organization
        \item identify products through barcode scanning
    \end{itemize}

    Each of those features transforms visual data into auditory information, helping visually impaired users interact with their environment. AIris' feedback is immediate and can improve the users' independence, helping with many daily tasks. Examples of the user-system interaction can be seen in Figure \ref{fig:img1} and Figure \ref{fig:img2}.

To measure AIris performance, we can note the machine learning models' accuracy in popular datasets \ref{table:ml_models_iris}. The high performance metrics and relatively low response times ensure a natural interaction with the aid.

\begin{table}[h]
\centering
\small 
\renewcommand{\arraystretch}{1.5} 
\begin{tabular}{|p{0.35\columnwidth}|p{0.35\columnwidth}|p{0.15\columnwidth}|}
\hline
\textbf{Model/Library}             & \textbf{Performance} & \textbf{Response Time} \\ \hline
\textbf{Face Recognition} - face\_recognition (ResNet-34)      & ~99.38\% accuracy on LFW dataset \cite{LFWTech}    &   50ms         \\ \hline
\textbf{Object Detection} - YOLO (You Only Look Once)          & ~63.4\% mean Average Precision on COCO dataset \cite{COCO}   & 150ms       \\ \hline
\textbf{Speech to Text} - Google's Speech Recognition       & ~95\% accuracy for English         &      200ms  \\ \hline
\textbf{Scene Description} - MAX Image Caption Generator       & ~27\% BLEU \cite{bleu} score on Flickr8k dataset \cite{flickr8k} & 150ms\\ \hline
\end{tabular}
\caption{Summary of Machine Learning Models used in AIris}
\label{table:ml_models_iris}
\end{table}

 AIris, in its current prototype form, has demonstrated significant potential in aiding visually impaired users. AIris modules have been tested in an unofficial trial to evaluate the practical applicability and user experience of AIris in real-world scenarios.

The trial involved three visually impaired individuals using AIris in a controlled environment. The participants were asked to engage in various tasks such as paying for something, identifying objects, reading text, and recognizing faces. Their feedback was collected to measure the effectiveness of AIris and identify areas for improvement.

The device's functionalities have been positively received in the trial as useful tools assisting  in  daily activities. These features offer practical support, enhancing the users' ability to interact with their environment more effectively. The presence of AIris itself has been found to provide a sense of reassurance to its users, while the real-time auditory feedback contributes to a sense of security and independence.

Through this informal trial, certain areas for improvement have been identified, which are crucial for enhancing the overall user experience and effectiveness of the device. One of the primary concerns noted is the latency in response times. This delay between user-command and system-feedback can be a significant setback, especially in scenarios where timely information is essential. Reducing this latency is critical for ensuring that AIris can be reliably used in dynamic and potentially fast-paced environments, like everyday activities. Additionally, the ergonomic design of the wearable device has been highlighted as an area needing attention. Prolonged use of the device has been reported to cause discomfort for some users, indicating a need for a more user-friendly design that can be comfortably worn for extended periods. This includes considerations for weight distribution, fit, and the overall comfort of the eyewear. The customization of user experience is another aspect that requires improvement. Users have expressed a desire for more personalized settings, such as adjustable speech output and tailored information delivery. Meeting these needs would significantly enhance user interaction, allowing for a more individualized and satisfying experience with AIris. Furthermore, while the accuracy of the machine learning models is generally high, there is always room for enhancement, especially in terms of adapting to varying environmental conditions and diverse user requirements.

  The current embodiment of AIris, while functional, remains in a prototypical phase. The technical implementation detailed herein reflects a preliminary but promising integration of sophisticated components and innovative software architectures designed to aid visually impaired individuals.
  
 AIris doesn't require a lot of resources since heavy machine learning models are running on the cloud. The system is flexible, updateable, and upgradeable. It can incorporate new features and functionalities, like new state-of-the-art machine learning models. Also, by situating the microcomputer in the user's pocket, we effectively distribute the system's weight, making the eyewear feel as close to ordinary glasses as possible.  Finally, both hardware and software can be personalized, from using different cameras and designing different styles of glasses to creating custom personalized features according to each user's needs. 

The technical overview of AIris's implementation emphasizes that while the current prototype can showcase significant potential, there is a clear roadmap for advancement to realize a fully refined assistive device.

Incorporating advanced AI technologies like Large Language Models (LLMs), Large Vision-Language Models, and Generative AI is set to significantly enhance assistive devices like AIris. These technologies offer an improved understanding of text and images, leading to more accurate and context-aware user interactions. They also enable more personalized responses, which can greatly improve the user experience. Our next step is to integrate these AI advancements into AIris, aiming to improve its current capabilities and introduce new features. This will help us better serve the needs of visually impaired individuals, offering them more independence and a richer interaction with their surroundings.
\section{CONCLUSIONS}
    In conclusion, AIris represents a transformative step in assistive technology for the visually impaired. This AI-powered wearable device, through its integration of advanced computer vision algorithms and natural language processing, offers great spatial awareness and interaction capabilities while maintaining an easy-to-upgrade user-friendly concept. Its innovative design, combining camera-equipped eyewear with auditory feedback, provides users with real-time environmental descriptions and object recognition, significantly enhancing their independence and interaction with the world.

    This technology can be used as a standalone solution or integrated with existing assistive technologies for the visually impaired.  Potential next steps involve conducting more thorough studies to evaluate AIris's performance on daily tasks, both qualitatively and quantitatively. Further enhancements to AIris functionalities, as well as refinements to existing features, may be pursued through a participatory design approach. In this process, collaboration with the visually impaired community is encouraged to actively co-create future projects.

\printbibliography

\end{document}